\documentclass[preprint,12pt]{elsarticle}
\usepackage{amssymb}
\usepackage{amsmath}
\usepackage{graphicx}
\usepackage{subfigure}
\journal{Optics Communications}

\begin{document}

\begin{frontmatter}
\title{Spatially-temporal dynamics of a passively Q-switched Raman-active solid-state oscillator}

\author[tuw]{V.L.Kalashnikov\corref{cor1}\fnref{akn}}
\ead{kalashnikov@tuwien.ac.at}
\ead[url]{http://info.tuwien.ac.at/kalashnikov}
\author[ilc]{A.M.Malyarevich}
\author[ilc]{K.V.Yumashev}
\cortext[cor1]{Corresponding author} \fntext[akn]{Author
acknowledges the support from the Austrian Fonds zur Forderung der
wissenschaftlichen Forschung (project P20293).}
\address[tuw]{Institut f\"{u}r Photonik, TU Wien, Gusshausstr. 27/387, A-1040 Vienna, Austria}
\address[ilc]{Institute for Optical Materials and Technologies, 65 Nezavisimosti Ave., Bldg. 17, 220013 Minsk, Belarus }

\begin{abstract}
The spatially-temporal model of an all-solid-state passively
Q-switched oscillator with an active medium providing the stimulated
Raman scattering is presented. The model does not presume a Gaussian
shape of the cylindrically symmetric modes at both fundamental and
Stokes wavelengths. It is found, that the highly-nontrivial
spatially-temporal dynamics can be regularized by the optimal choice
of the oscillator parameters, viz. initial transmission of a
saturable absorber, curvature of a spherical mirror, and output
mirror transmission at the fundamental and Stokes wavelengths. As a
result, the pulse can be substantially temporally squeezed and
spatially broadened at both fundamental and Stokes wavelengths.
\end{abstract}


\end{frontmatter}

\section{Introduction}
\label{intro}

Solid-state Q-switched oscillators allowing nano- and sub-nanosecond
pulsing find applications in a lot of areas including medicine,
spectroscopy, environment monitoring, etc. Passive Q-switching based
on the use of both semiconductor \cite{keller,yu} and crystalline
\cite{m} saturable absorbers is of particular interest due to
compactness, simplicity, high damage threshold, and diode-pumping
ability of an oscillator.

Among the active media allowing diode-pumped Q-switching,
KY(WO$_4$)$_2$ (KYW) and KGd(WO$_4$)$_2$ (KGW) crystals doped by
Yb$^{3+}$ and Nd$^{3+}$ ions are known as the materials providing an
efficient stimulated Raman scattering (SRS)
\cite{eich,kuleshov1,kuleshov2,grab1,grab2}. As a result, there is
possible a high-efficient self-frequency shift and a simultaneous
two-wavelengths pulsing (e.g., at 1.35 and 1.54 $\mu$m for a Nd:KGW
active medium) directly from an oscillator.

Theoretical studies of a Q-switched oscillator with the intra-cavity
SRS have been based on the well-established rate-equations approach
\cite{jap,kalash,loiko}. The oscillator parameters providing a pulse
width minimization \cite{kalash} and an output energy maximization
\cite{loiko} at both fundamental and Stokes wavelengths have been
defined. Simultaneously, it has been found that the spatial
structure of a laser field is strongly affected by the SRS
\cite{jap}. This means that the mode transformation has to be taken
into account along with the temporal evolution of fields and
populations inside an active medium and a saturable absorber. Such a
model has been developed to a Gaussian mode approximation for both
fundamental and Stokes fields \cite{spatial}. This has allowed
defining the optimal values of the Raman gain and the ratio of pump
and laser beams, which provide a single pulse operation of
oscillator with the most efficient SRS.

Nevertheless, a transient character of Q-switching can prevent from
the CW mode formation and disturb substantially the spatial
structure of a laser field. This requires to take into account the
spatial dynamics on a par with the temporal one.

In this work we present the analysis of spatially-temporal dynamics
of a passively Q-switched Raman-active oscillator. The analysis is
based on a rate-equations approach but without imposing a limitation
on the shape of cylindrically symmetric transverse distribution of a
field. The results demonstrate that there exist both spatial
extra-broadening and squeezing as well as transition between
non-Gaussian and Gaussian spatial profiles in dependence on the
oscillator parameters (viz. output mirror transmission at the
fundamental and Stokes wavelengths, initial transmission of a
saturable absorber and curvature of a spherical mirror). The SRS
contribution controlled by the optimized sets of oscillator
parameters allows the regularization of the spatially-temporal
structure of a field and the substantial pulse shortening.

\section{Numerical procedure}
\label{numer}

The oscillator under consideration consists of the flat (output) and
spherical mirrors. The active medium (Nd$^{3+}$:KGW) is placed at
the resonator center, and the saturable absorber (V$^{3+}$:YAG) is
placed on the output mirror.

The model is based on a direct generalization of that presented in
\cite{kalash,spatial}. It is supposed that the time-dependent ($t$
is the time, which is periodical with the cavity period $T_{cav}$:
${t\in\left[ {0,T_{cav} } \right]} $) fields $a_{f,s}$ (indexes $f$
and $s$ correspond to the fundamental and Stokes fields,
respectively) have the radially-symmetric transverse spatial
distributions ($ {r\in\left] {0,R} \right]} $ is the radial
coordinate).

Inside an active medium, the dynamics is modeled on basis of the
split-step Hankel's method. That is the active medium volume is
considered to be divided into 10 transverse slices with the
thickness $\Delta z$ ($z$ is the longitudinal coordinate), in which
the dynamics within the time-domain and the domain of spatial
frequencies ($\nu \equiv c/2\pi R$, $c$ is the velocity of light,
$R=0.5\div1.5$ cm) \cite{siegman} is evolved step-by-step:

\begin{eqnarray}\label{am}
\left. \begin{gathered}
  \dot a_f \left( {z,r,t} \right) = \frac{{\Delta z}}
{2 T_{cav}}a_f \left( {z,r,t} \right)\left[ {g\left( {r,t} \right) - g_s \left| {a_s \left( {z,r,t} \right)} \right|^2 } \right] \hfill \\
  \dot a_s \left( {z,r,t} \right) = \frac{{\Delta z}}
{2 T_{cav}}a_s \left( {z,r,t} \right)g_s \left| {a_f \left( {z,r,t} \right)} \right|^2  \hfill \\
  \dot g\left( {r,t} \right) =  - \frac{{\sigma_g \lambda_f}}
{h c} g\left( {r,t} \right)\left| {a_f \left( {z,r,t} \right)} \right|^2  \hfill \\
\end{gathered}  \right\} \otimes\\ \nonumber \otimes \tilde a_{f,s}\left( {z+\Delta z,\nu,t} \right)  = \tilde a_{f,s}\left( {z,\nu,t} \right) \exp \left( { - ik_{f,s} \Delta z + \frac{i}
{2}k_{f,s} \Delta z\lambda _{f,s}^2 \nu ^2 } \right).
\end{eqnarray}

\noindent We shall suppose, that the field intensities $\left|
a_{f,s} \right|^2$ are normalized to $h c/\lambda_s \sigma_g
T_{cav}$ ($h$ is the Planck constant; $\lambda_f=$1.35 $\mu$m and
$\lambda_s=$1.54 $\mu$m are the fundamental and Stokes wavelengths,
respectively; $\sigma_g=$0.76$\times10^{-19}$ cm$^2$ is the gain
cross-section, $T_{cav}=$0.8 ns), and the time is normalized to
$T_{cav}$. The gain coefficient is $g$ and its initial value equals
to the threshold one: $(\ln{1/T_0^2}+\ln{1/\rho_f}+l)/2 L_g$. In the
last expression, the varied values $T_0$ and $\rho_f$ correspond,
respectively, to the initial transmission of saturable absorber and
the reflection of output mirror at the fundamental wavelength;
$l=0.05$ is the unsaturable net-loss coefficient. $L_g=$5 cm is the
active medium length. The stimulated Raman scattering inside an
active medium is described by the coefficient $g_s$ and its
dimensional value amounts to 6 cm/GW. It is assumed that there
exists no anti-Stokes and higher-order Stokes scattering as well as
that the gain is saturable by only fundamental field. Also, the
initial gain distribution is assumed to be spatially homogeneous.

The field representations $\tilde a_{f,s}$ within the spatial
frequency domain are obtained by the means of the fast Hankel's
transformation \cite{siegman,hankel}. $k_{f,s}$ are the wave-numbers
corresponding to the fundamental and Stokes fields, respectively.

A similar procedure describes the dynamics inside a saturable
absorber:

\begin{eqnarray} \label{absorber}
\left. \begin{gathered}
  \dot a_f \left( {z,r,t} \right) =  - \frac{{\Delta z}}
{2 T_{cav}}a_f \left( {z,r,t} \right)\left[ {n\left( {r,t} \right) - \sigma _{esa} \left( {N - n\left( {r,t} \right)} \right)} \right] \hfill \\
  \dot a_s \left( {z,r,t} \right) =  - \frac{{\Delta z}}
{2 T_{cav}}a_s \left( {z,r,t} \right)n\left( {r,t} \right) \hfill \\
  \dot n\left( {r,t} \right) =  - \frac{{\sigma_f \lambda_f}}
{h c} n\left( {r,t} \right)\left| {a_f \left( {z,r,t} \right)}
\right|^2  - \frac{{\sigma_s \lambda_f}} {h c} n\left( {r,t}
\right)\left| {a_s \left( {z,r,t} \right)} \right|^2  + \frac{{N -
n\left( {r,t} \right)}}
{{T_r }} \hfill \\
\end{gathered}  \right\} \otimes\\ \nonumber
 \otimes \tilde a_{f,s}\left( {z+\Delta z,\nu,t} \right)  = \tilde a_{f,s}\left( {z,\nu,t} \right) \exp \left( { - ik_{f,s} \Delta z
+ \frac{i} {2}k_{f,s} \Delta z\lambda _{f,s}^2 \nu ^2 } \right)
\end{eqnarray}

\noindent Here, $n$ is the loss coefficient and its initial value
amounts to $N=(\ln{1/T_0^2})/2 L_a$ ($L_a$=0.1 cm is the saturable
absorber thickness). The loss is saturable by both fundamental and
Stokes fields ($\sigma_f=95$, $\sigma_s=4$, when the normalization
to $\sigma_g$ is presumed). Excited-state absorption is taken into
account for the fundamental field ($\sigma_{esa}=$0.1). $T_r=26$ is
the loss relaxation time (the normalization to $T_{cav}$ is
presumed).

Free propagations between the active medium and the absorber as well
as between the active medium and the spherical mirror are considered
in the frequency domain:

\begin{equation}\label{free}
\tilde a_{f,s}\left( {z+L,\nu,t} \right)  = \tilde a_{f,s}\left(
{z,\nu,t} \right) \exp \left( { - ik_{f,s} L + \frac{i} {2}k_{f,s}
L\lambda _{f,s}^2 \nu ^2 } \right),
\end{equation}

\noindent where $L=5$ cm is the propagation length.

The reflection from a spherical mirror obeys:

\begin{equation}\label{mirror}
a_{f,s} \left( {z,r,t} \right) = a_{f,s} \left( {z,r,t} \right)\exp
\left( {{{ik_{f,s} r^2 } \mathord{\left/
 {\vphantom {{ik_{f,s} r^2 } {R_M }}} \right.
 \kern-\nulldelimiterspace} {R_M }}} \right)
\end{equation}

\noindent where $R_M$ is the variable radius of the mirror
curvature.

The loss on an output mirror as well as the net unsaturable loss are
taken into account by means of the following mapping:

\begin{equation}\label{mirror}
a_{f,s} \left( {z,r,t} \right) = a_{f,s} \left( {z,r,t} \right)\exp
\left[ -0.5 (l+\ln(1 /\rho_{f,s})) \right],
\end{equation}

\noindent where $\rho_s$ is the output mirror reflection at the
Stokes wavelength.

In the simulations, the time window is approximated by 100 points
and the window of transverse coordinate is approximated by
2000$\div$6000 points in dependence on the $R$ value, which varies
from 0.5 to 1.5 cm in order to exclude the boundary effects. The
solution in the time domain is evaluated using a fourth-order
Runge-Kutta method. The initial dimensionless fields have the
Gaussian spatial profile with the size, which equals to that of the
fundamental CW mode at $\lambda_f$. The initial dimensionless
amplitudes $a_f$ and $a_s$ amount to 10$^{-2}$ and 10$^{-10}$,
respectively.

\section{Spatially-temporal structure of laser field from a Q-switched oscillator}
\label{results}

The simulations based on the model described in the previous Section
demonstrate that the spatially-temporal structure of laser field
from a Q-switched oscillator depends non-trivially on the oscillator
parameters, which effect on the field dynamics at both fundamental
and Stokes wavelengths.

\begin{figure}
\centering
  \includegraphics[width=13cm]{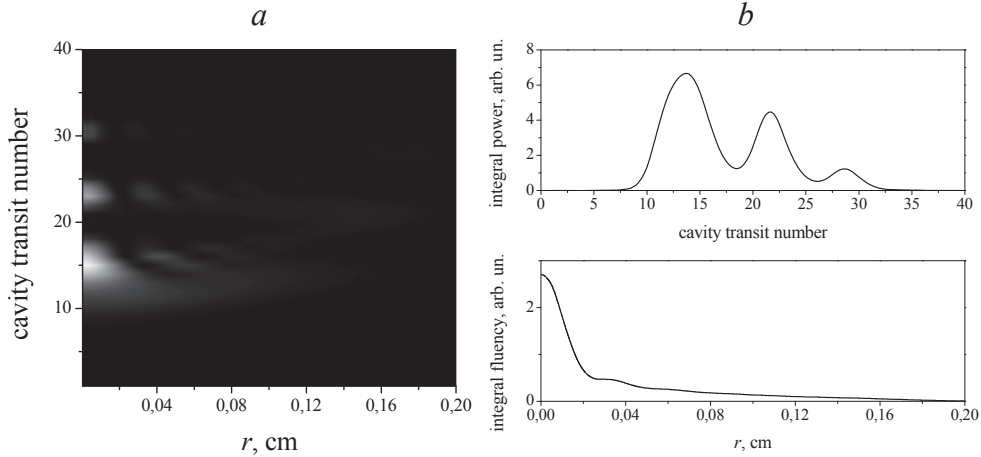}\\
  \caption{\emph{a}: Contour plot of the output intensity at the fundamental wavelength from a Q-switched oscillator
  with $T_0=$0.3, $\rho_f=$0.7, $\rho_s=$0, $R_M=$300 cm. \emph{b}: Output intensity integrated over $r$
  (integral power; upper plot) or $t$ (integral fluency; lower plot).}\label{fig1}
\end{figure}

\begin{figure}
\centering
  \includegraphics[width=13cm]{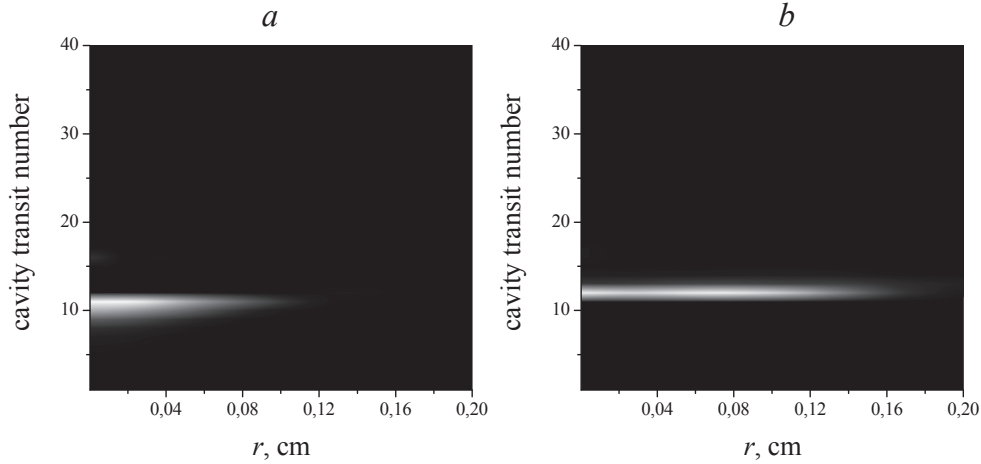}\\
  \caption{Contour plots of the output intensity at the fundamental (\emph{a}) and Stokes (\emph{b}) wavelengths.
  $T_0=$0.3, $\rho_f=$0.7, $\rho_s=$0.7,
  $R_M=$300 cm.}\label{fig2}
\end{figure}

Fig. \ref{fig1}, \emph{a} shows the contour plot of the output
intensity at the fundamental wavelength in absence of an effective
SRS in an oscillator ($\rho_s=$0). The intensities are averaged over
the cavity period in Figs. \ref{fig1}, \emph{a}; \ref{fig2},
\ref{fig4}, \ref{fig5}, \ref{fig7}, and \ref{fig8}. Both low initial
transmission of a saturable absorber and comparatively small radius
of curvature of a spherical mirror initiate the multiple pulse
dynamics, which is clearly visible in Fig. \ref{fig1}, \emph{b}
(upper plot) showing the output intensity integrated over the
transverse spatial coordinate. The greatest pulse is comparatively
long ($\approx$5 ns) and the narrow ``mode'' (the radius equals to
$\approx$130 $\mu$m) has an elongate wing (lower plot). As will be
shown, a non-Gaussian transverse shape is typical for the regime
under consideration (see \cite{jap}).

\begin{figure}
\centering
  \includegraphics[width=8cm]{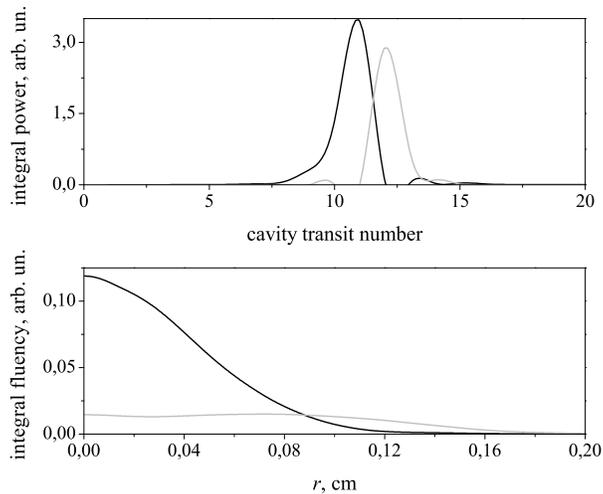}\\
  \caption{Output integral power (upper plot) and integral fluency (lower plot)
  at the fundamental (black curves) and Stokes (gray curves) wavelengths.
  Parameters correspond to Fig. \ref{fig2}.}\label{fig3}
\end{figure}

\begin{figure}
\centering
  \includegraphics[width=13cm]{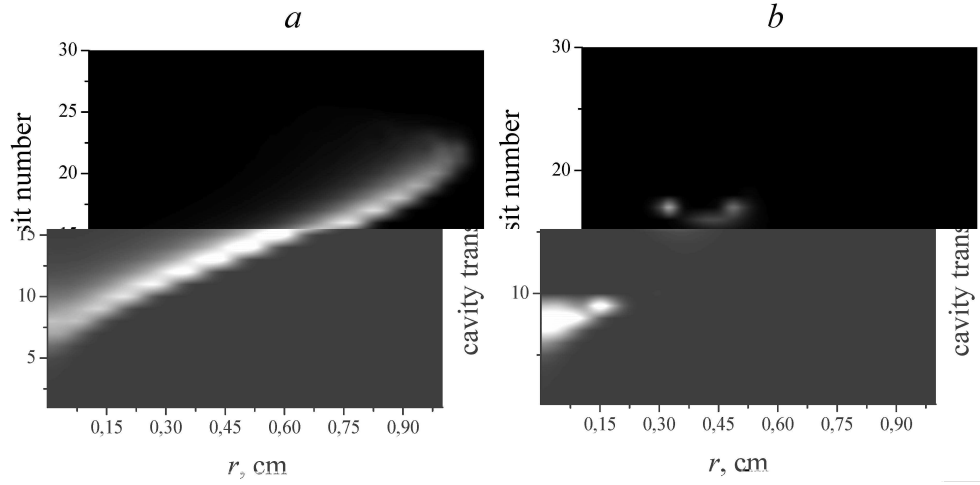}\\
  \caption{Contour plots of the output intensity at the fundamental wavelength. $T_0=$0.3, $\rho_f=$0.5; $\rho_s=$0 (\emph{a}), 0.7
  (\emph{b});
  $R_M=$2000 cm.}\label{fig4}
\end{figure}

\begin{figure}
\centering
  \includegraphics[width=14cm]{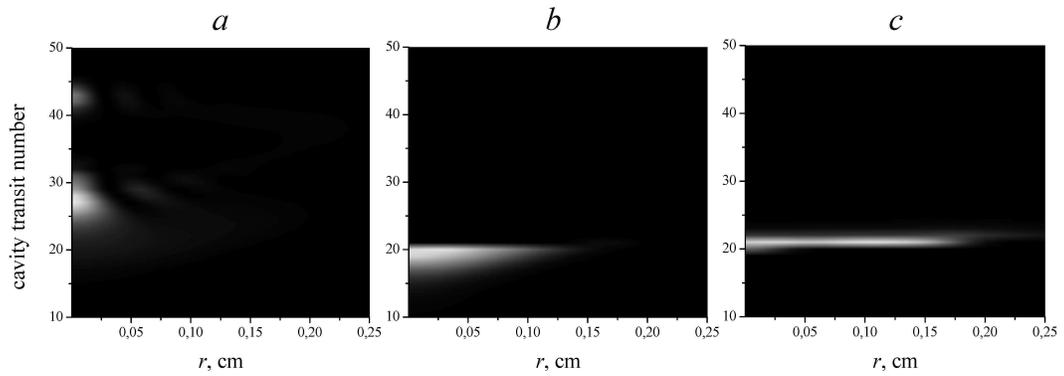}\\
  \caption{Contour plots of the output intensity at the fundamental (\emph{a}, \emph{b}) and Stokes (\emph{c}) wavelengths.
  $T_0=$0.5, $\rho_f=$0.9; $\rho_s=$0 (\emph{a}), 0.9
  (\emph{b}, \emph{c}); $R_M=$1000 cm.}\label{fig5}
\end{figure}

When a field at the Stokes wavelength is locked inside an oscillator
owing to non-zero $\rho_s$, the dynamics changes. For the case
presented in Fig. \ref{fig1}, the multipulsing disappears (Figs.
\ref{fig2}, \emph{a} and \ref{fig3}) and the pulse at the
fundamental wavelength shortens substantially ($\approx$1.2 ns, see
Fig. \ref{fig3}, upper plot) \cite{kalash}. The mode at the
fundamental wavelength widens and becomes Gaussian (Fig. \ref{fig3},
lower plot; see also \cite{jap}). Simultaneously, the short pulse at
the Stokes wavelength develops (Figs. \ref{fig2}, \emph{b} and
\ref{fig3}). Such a pulse is located on the tale of the fundamental
pulse (Fig. \ref{fig3}, gray curve in upper plot) and has a broad
trapezoidal transverse profile (the mode radius reaches $\approx$1.4
mm in Fig. \ref{fig3}, gray curve in lower plot).

Thus, the transverse distribution excesses substantially the CW mode
size, which equals to $\approx$265 $\mu$m for the spherical mirror
under consideration. Such a spatial extra-broadening can be
attributed to the saturable absorber saturation, which results in an
excitation of high-order spatial frequencies.

A spatial extra-broadening is possible at the fundamental
wavelength, as well. Fig. \ref{fig4}, \emph{a} demonstrates such an
extra-broadening (up to $\approx$0.9 cm). The field covers almost a
whole active crystal radius. It should be noted, that the output CW
mode radius equals to $\approx$440 $\mu$m for the spherical mirror
under consideration. The Q-switching ``mode'' has a shape of a
dilative ring, which rises and then disappears (after $\approx$12
ns) with a gain depletion. In the presence of SRS, such a ``mode''
is suppressed owing to an inhomogeneous ``collapsing''
spatially-temporal behavior at both fundamental and Stokes
wavelengths (Fig. \ref{fig4}, \emph{b} and \cite{kalash2}).

\begin{figure}
\centering
  \includegraphics[width=8cm]{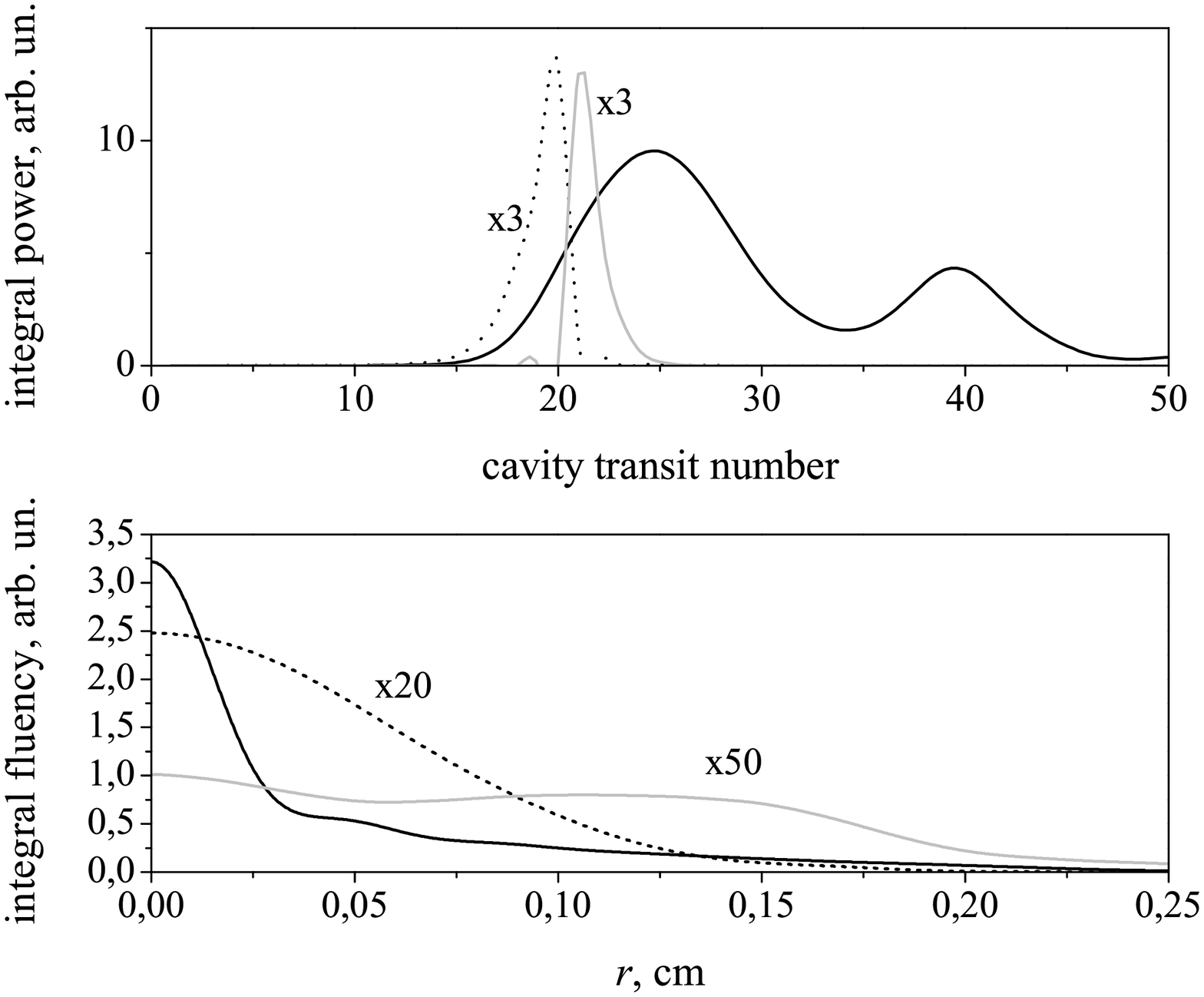}\\
  \caption{Output integral power (upper plot) and integral fluency (lower plot)
  at the fundamental (black solid and dotted curves) and Stokes (gray curves) wavelengthes.
  $\rho_s=$0 (black solid curves) and 0.9  (black dotted and gray curves). Another parameters correspond to Fig. \ref{fig5}.
  Dotted and gray curves are
  vertically rescaled for convenience.}\label{fig6}
\end{figure}

\begin{figure}
\centering \subfigure{\includegraphics[width=7cm]{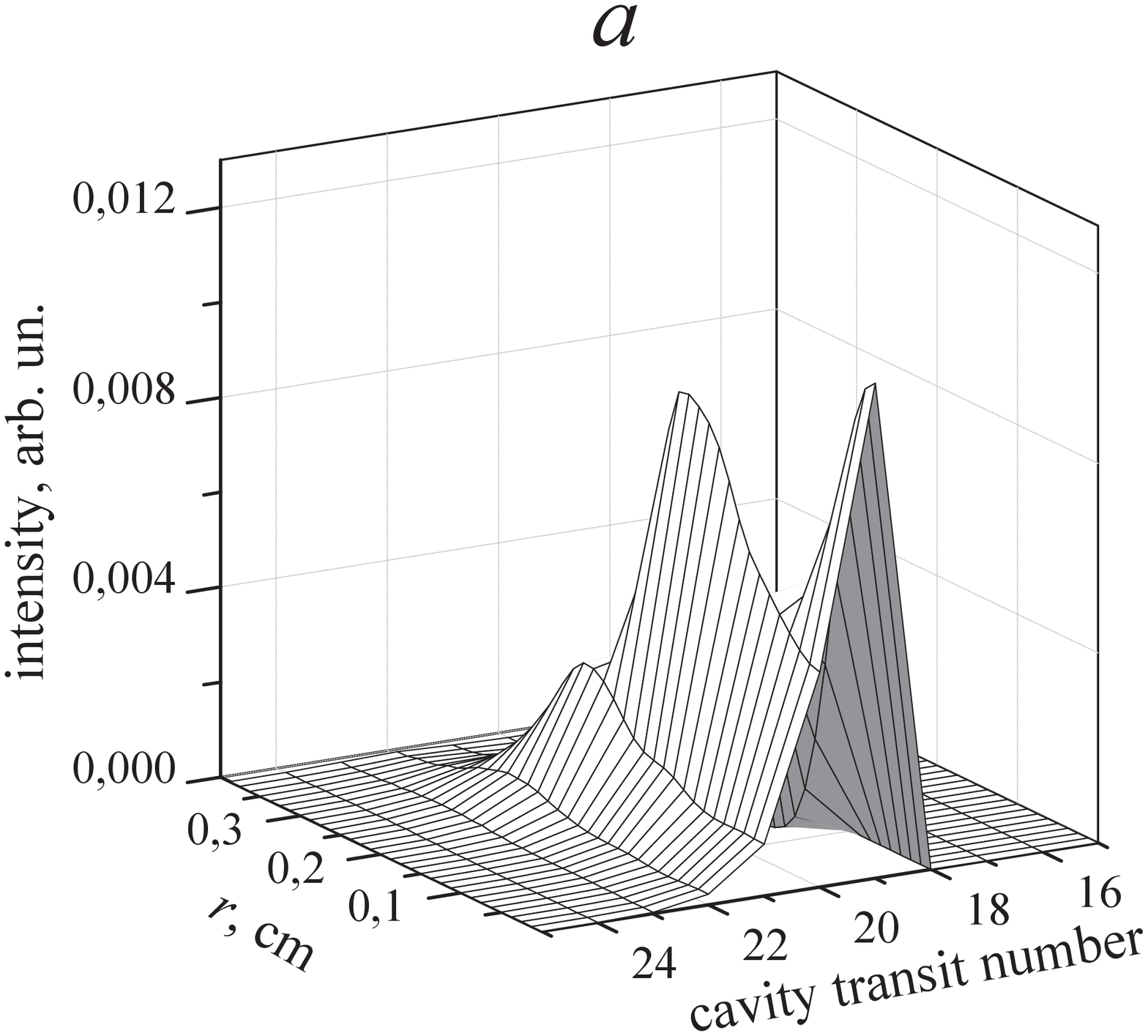}
\includegraphics[width=7cm]{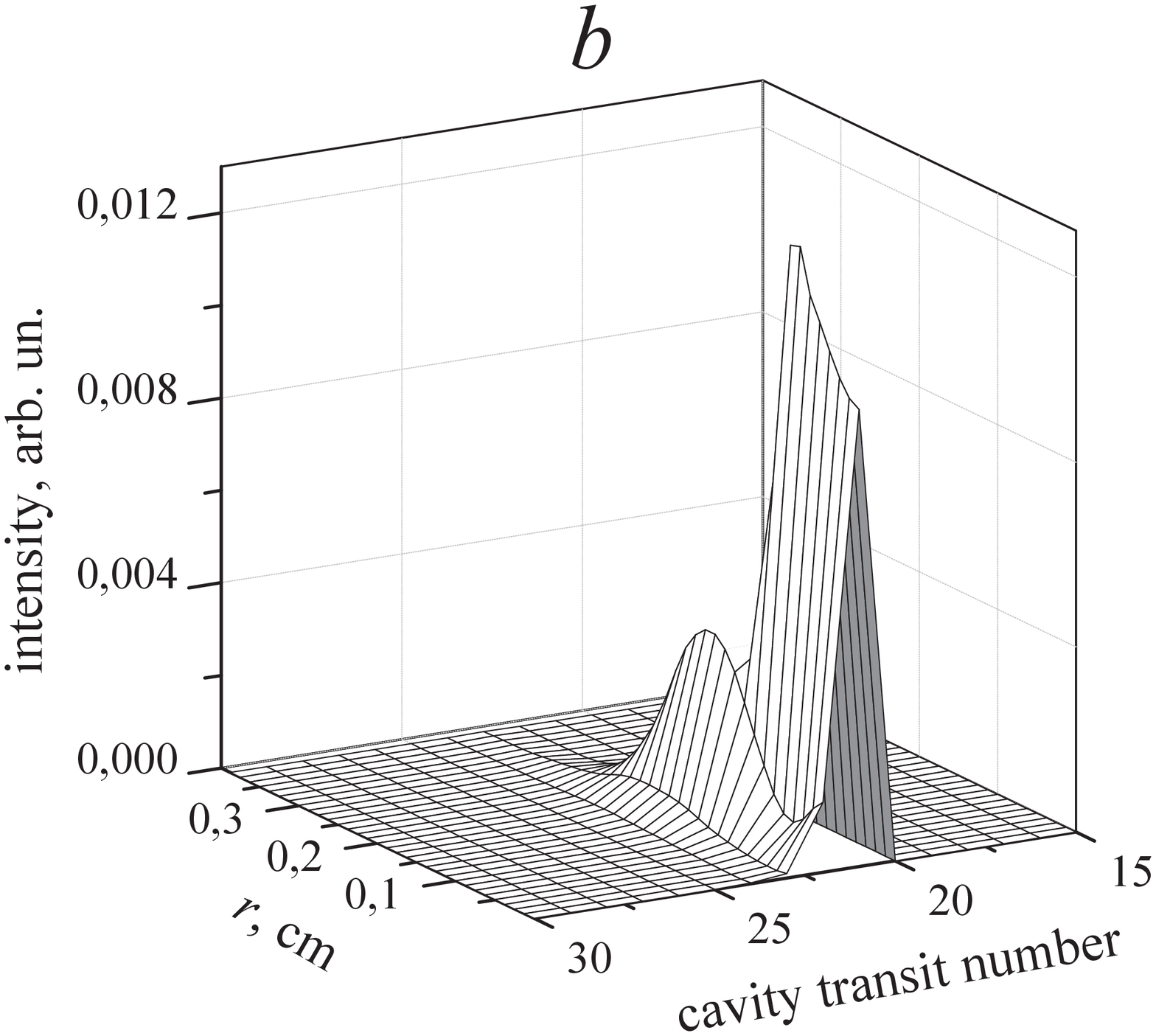}}
  \caption{Stokes intensity profiles for (\emph{a}): $R_M$=2000 cm,
  $\rho_f$=0.7, $\rho_s$=0.7, and (\textbf{b}): $R_M$=1000 cm,
  $\rho_f$=0.9, $\rho_s$=0.5. $T_0$=0.5.}\label{fig7}
\end{figure}

\begin{figure}
\centering \includegraphics[width=7cm]{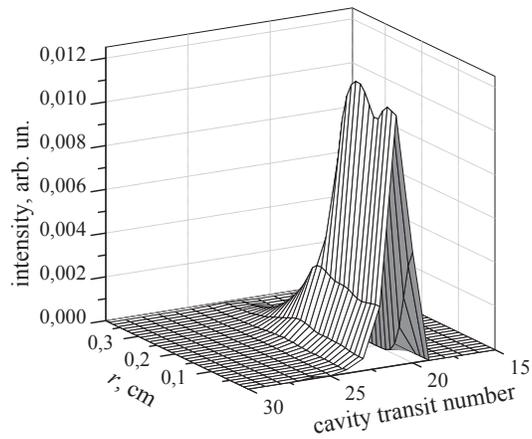}
  \caption{Stokes intensity profile for $R_M$=1000 cm,
  $\rho_f$=0.9, $\rho_s$=0.9, and $T_0$=0.5.}\label{fig8}
\end{figure}

An effect of the SRS on the Q-switching dynamics allows controlling
the temporal and spatial profiles of a pulse at both fundamental and
Stokes wavelengths. An imbalance between the loss and gain
saturation at the fundamental wavelength causing the multiple
pulsing can be compensated by the nonlinear loss due to SRS. As a
result, the well-shaped and substantially shortened \cite{kalash}
pulses with the broad spatial profiles at both wavelengths appear
(Figs. \ref{fig5},\ref{fig6}).

When the oscillator parameters are not optimized, the
spatially-temporal structure of the Stokes field can be complicated.
For instance, the radial position of the SRS spike can depend on the
time (Fig. \ref{fig7}, \emph{a}, where the SRS spikes move off the
optical axis with the pulse evolution; this effect has been observed
experimentally in \cite{dash}). A similar situation is possible also
at the fundamental wavelength (Fig. \ref{fig4}, \emph{a}). Even
through the SRS is well-synchronized (i.e. the SRS appears
synchronously at the different radial positions), the mode profile
is inhomogeneous, as a rule (Fig. \ref{fig7}, \emph{b}).

The oscillator parameters optimization allows regularizing the
spatially-temporal dynamics at both fundamental and Stokes
wavelengths. In particular, the spatial profile of the Stokes field
becomes homogeneous though non-Gaussian (Fig. \ref{fig8}). The
numerical analysis reveals, that there exist some optimal sets of
parameters (see Table \ref{t1}), in particular the optimal curvature
of a spherical mirror ($\approx$1000 cm in the case under
consideration), which provide the regular pulses within a broadest
range of parameters (viz. $\rho_f$, $\rho_s$, and $T_0$). The
initial transmission $T_0>$0.5 results in a sufficiently strong
fundamental field, that causes an efficient SRS. The reflectivity at
the Stokes wavelengths has to be sufficiently high ($>$0.7) to
provide an efficient conversion from the fundamental field to the
Stokes one.

The resulting pulses at both wavelength are shortened in comparison
with a single-wavelength regime and have the smooth and broad
transverse spatial profiles. The last property allows using a larger
volume of an active crystal. The ``mode'' shape is close to the
Gaussian one for the fundamental field and to the trapezoidal one
for the Stokes field.

\begin{table}
  \centering
  \caption{Oscillator parameters providing the regular spatial-temporal pulse profiles
  at both fundamental and Stokes wavelengths.}\label{t1}
  \begin{tabular}{|c|c|c|c|}\hline
$R_M$, cm& $T_0$ & $\rho_f$ & $\rho_s$ \\ \hline\hline 300& 0.3 --
0.5 & 0.7 -- 1 & 0.7 -- 0.9 \\ \hline 1000& 0.3 -- 0.5 & 0.5 -- 1 &
0.7 -- 0.9 \\ \hline 2000& 0.5 & 0.9 -- 1 & 0.9\\ \hline
\end{tabular}

\end{table}

\section{Conclusion}
\label{concl}

The model of a Q-switched Nd:KGW oscillator with a V:YAG saturable
absorber, which takes into account a transverse spatial field
distribution at both fundamental and Stokes wavelengths, has been
developed. The numerical analysis has demonstrated that an imbalance
between the loss and gain saturation causes not only a multiple
pulsing but an aberration of an oscillator ``mode''. When the
initial transmission of a saturable absorber is sufficiently low,
the spatial extra-broadening of a field develops so that a ``mode''
covers almost whole volume of an active crystal.

Manipulations with the beam size on a saturable absorber (by means
of the change of a spherical mirror curvature), the initial
transmission of an absorber, and the output mirror reflectivity at
both fundamental and Stokes wavelengths allow controlling the
spatially-temporal profiles of the output pulses. To obtain a single
pulsing at two-wavelengths with a broad and homogeneous spatial
distribution, the initial transmission of the saturable absorber has
to be about of 0.5, the reflectivity of the output mirror at the
fundamental wavelength has to approach 1 (it can range within 0.5 --
1 for the optimal curvature of the spherical mirror, which is of
$\approx$1 m in the case under consideration), and the reflectivity
at the Stokes wavelength has to be about of 0.9. The SRS causes a
temporal squeezing of the pulses at both wavelengths, whereas the
spatial profiles are greatly stretched in comparison with the
CW-mode size. When the oscillator parameters are optimized, the
transverse spatial profiles are near-Gaussian at the fundamental
wavelength and trapezoidal at the Stokes wavelength.

The obtained results are of interest for development of the
passively Q-switched Raman-active oscillators producing the pulses
with durations of about of few nanoseconds and possessing the smooth
transverse spatial profiles covering a considerable part of an
active crystal.

\end{document}